\documentclass[10pt,conference]{IEEEtran}
\usepackage[top=0.7in, bottom=0.98in, left=0.7in, right=0.7in]{geometry}
\IEEEoverridecommandlockouts
\usepackage{cite}
\usepackage{subfig}
\usepackage{amsmath,amssymb,amsfonts}
\usepackage{algorithmic}
\usepackage{graphicx}
\usepackage{textcomp}
\usepackage{xcolor}
\usepackage{graphicx}
\usepackage{float}
\usepackage{bm}
\newtheorem{remark}{\rm{\textbf{Remark}}}
\DeclareSubrefFormat{parens}{#1(#2)}
\def\BibTeX{{\rm B\kern-.05em{\sc i\kern-.025em b}\kern-.08em
    T\kern-.1667em\lower.7ex\hbox{E}\kern-.125emX}}
\usepackage{cmap} 
\usepackage[T1]{fontenc}
\usepackage{times} 
\usepackage[utf8]{inputenc}

\makeatother
\begin{document}

\title{Enhancing Environment Generalizability for Deep Learning-Based CSI Feedback
}
\vspace{-20pt}
\author{\IEEEauthorblockN{Haoyu Wang\textsuperscript{1}, Shuangfeng Han\textsuperscript{2}, Xiaoyun Wang\textsuperscript{2}, and Zhi Sun\textsuperscript{1}}
\IEEEauthorblockA{\textsuperscript{1}Beijing National Research Center for Information Science and Technology (BNRist), \\Department of Electronic Engineering, Tsinghua University, Beijing 100084, China\\
\textsuperscript{2}China Mobile Research Institute, Beijing 100053, China.\\
\mbox{wanghy22@mails.tsinghua.edu.cn, \{hanshuangfeng,wangxiaoyun\}@chinamobile.com}, zhisun@ieee.org}}

\maketitle
\vspace{-20pt}
\begin{abstract}
Accurate and low-overhead channel state information (CSI) feedback is essential to boost the capacity of frequency division duplex (FDD) massive multiple-input multiple-output (MIMO) systems. Deep learning-based CSI feedback significantly outperforms conventional approaches. Nevertheless, current deep learning-based CSI feedback algorithms exhibit limited generalizability to unseen environments, which obviously increases the deployment cost. In this paper, we first model the distribution shift of CSI across different environments, which is composed of the distribution shift of multipath structure and a single-path. Then, EG-CsiNet is proposed as a novel CSI feedback learning framework to enhance environment-generalizability. Explicitly, EG-CsiNet comprises the modules of multipath decoupling and fine-grained alignment, which can address the distribution shift of multipath structure and a single path. Based on extensive simulations, the proposed EG-CsiNet can robustly enhance the generalizability in unseen environments compared to the state-of-the-art, especially in challenging conditions with a single source environment.
\end{abstract}

\section{Introduction}
Massive multiple-input multiple-output (MIMO) is a pivotal technology to enable high spectral efficiency in 5G and B5G wireless networks \cite{jsac_jin_2023_massive}. To leverage the large-scale antenna array in the massive MIMO system, precise knowledge of downlink channel state information (CSI) is essential to maximizing the performance of precoding and beamforming. In frequency division duplex (FDD) massive MIMO systems, the uplink and downlink channel is not reciprocal due to the gap between frequency bands. Thus, the acquisition of downlink CSI relies on the channel feedback procedure, which compresses the high-dimensional CSI at the user equipment (UE) and feeds the compressed codeword back to the base station (BS). Currently, codebook-based CSI feedback schemes are widely adopted in FDD massive MIMO systems \cite{3gpp.38.214}. However, the codebook-based feedback schemes cannot effectively model the complex correlations in the angular-delay domain of CSI, which exhibits relatively low compression capability \cite{icc_sang_2024_type2}. 

Deep learning exhibits powerful compress capability in modeling the correlations in CSI, which can enable low-overhead and accurate CSI feedback \cite{wcl_wen_2018_csinet,twc_guo_2020_csinetp, wcl_cui_2022_transnet}. Typically, the auto-encoder (AE) structure is widely adopted in current deep learning-based CSI feedback, which comprises an encoder module at the UE side and a decoder module at the BS side. To model the angular and delay domain correlations, different types of neural network (NN) structures have been proposed in the encoder and decoder modules, including the convolutional neural network (CNN)-based CsiNet/CsiNet+ \cite{wcl_wen_2018_csinet,twc_guo_2020_csinetp} and the transformer-based TransNet \cite{wcl_cui_2022_transnet}. 

Despite the powerful compression capability, current deep learning-based CSI feedback algorithms encounter the difficulties of generalizability, which has attracted the attention of both academia \cite{tcom_guo_2022_overview} and industry \cite{3gpp.38.843}. Explicitly, the deep learning-based CSI feedback algorithms are realized under in-distribution assumptions, where the test and training samples are drawn from the same distribution. However, in the practical deployment scenarios of massive MIMO systems, the distribution of CSI is determined by the electromagnetic wave propagation conditions in different environments, including the user distributions, BS location, layouts, and materials of objects in the wireless channel. Thus, the generalizability of the trained model cannot be guaranteed in an unseen environment distinct from the training environment. 

\begin{figure}[t]
        \centering
        \includegraphics[width=0.42\textwidth]{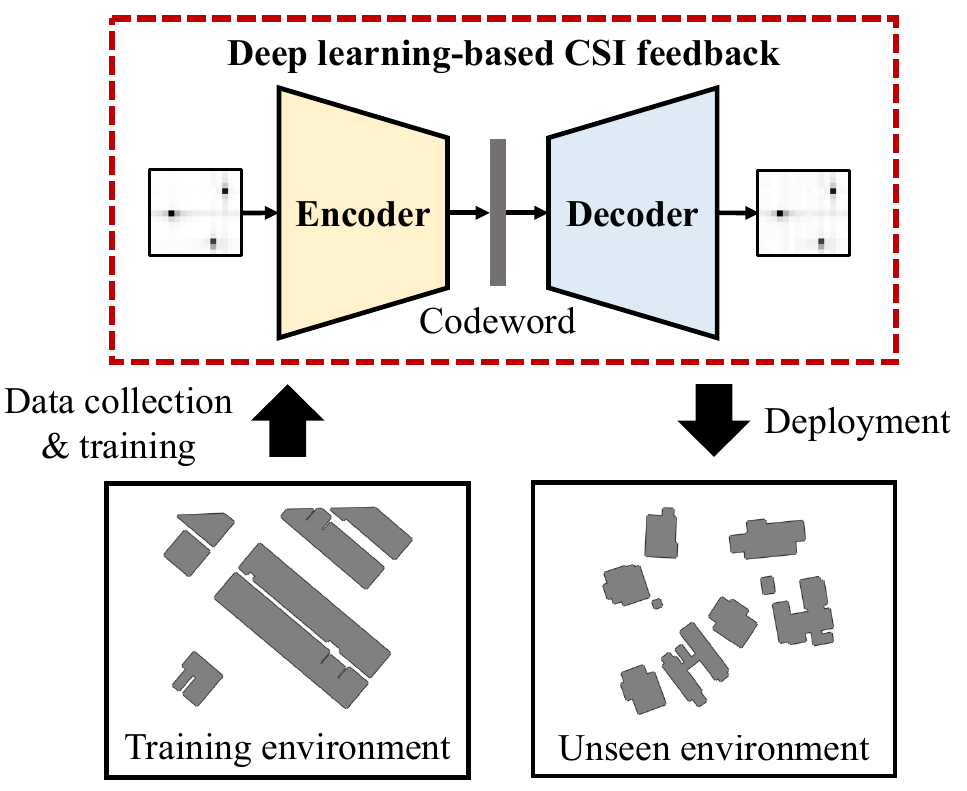}
    \captionsetup{font=footnotesize}
    \caption{Illustration of deep-learning based CSI feedback scheme with environment generalizability.}
    \label{fig: generalizability}
    \vspace{-20pt}
\end{figure}

As shown in Fig.~\ref{fig: generalizability}, environment-generalizable learning can guarantee the performance of CSI feedback models in unseen environments without additional training data, which can greatly reduce the deployment cost. Current state-of-the-art (SOTA) environment generalizable learning algorithms for CSI feedback include the dataset-mixing \cite{twc_jiang_2024_multidomain} and UniversalNet/UniversalNet+ \cite{wcnc_liu_2024_generalization}. In the dataset-mixing, the model is trained on a mixed dataset collected from different sources to enhance generalizability. However, no specialized generalization modules are designed, where the model generalizability is limited by the diversity of the mixed dataset. 
Meanwhile, the preparation of a large and diverse training dataset will incur significant time and labor costs in the real world. In UniversalNet/UniversalNet+, a benchmark CSI sample is created for both model training and evaluation, and the other samples are aligned with the benchmark via cross-correlation to reduce the inter-environment gap. Nevertheless, it is difficult to precisely align diverse CSI samples to a single benchmark, especially in complex multipath scenarios. 

To this end, we propose an efficient learning framework for CSI feedback with strong environment-generalizability in this paper, especially in complex multipath scenarios. Firstly, we model distribution shift CSI samples in different environments, including the distribution shift of the multipath structure and a single-path. Additionally, the distribution shift of a single-path in the angular-delay domain is investigated, including the shift of peak position, power leakage effect, and complex gain. Secondly, EG-CsiNet is proposed to address the distribution shift of CSI samples, which is composed of the multipath decoupling and fine-grained alignment. Motivated by our earlier work \cite{wang2025path}, each decoupled path component in EG-CsiNet is individually fed back to the BS, which can address the distribution shift of the multipath structure. Then, fine-grained alignment is proposed to address the distribution shift of the decoupled path component. Facilitate with the multipath decoupling and fine-grained alignment, model training, and inference of EG-CsiNet are also designed. Based on comprehensive simulation studies over an open-source dataset, the proposed EG-CsiNet can reduce the error of CSI feedback for more than 3.5 dB in unseen environments compared to the SOTA.

\textit{Notations:} $\mathbb{R}^{m\times n}$ and $\mathbb{C}^{m\times n}$ denote the real and complex spaces with dimension $m\times n$ and $\rm{j}=\sqrt{-1}$; $(\cdot)^{T}$, $\text{conj}(\cdot)$, and $(\cdot)^{H}$ denote the transpose, conjugate, and Hermitian transpose, respectively; $\otimes$ and $\odot$ stand for the Kronecker product and Hadamard product; $\Vert\mathbf{X}\Vert_{F}$ denotes the Frobenius norm of matrix $\mathbf{X}$ and $\Vert\mathbf{x}\Vert_{2}$ stands for Euclidean norm of vector $\mathbf{x}$; $[x]$ and $\lceil x\rceil$ denote the round and ceiling of real scalar $x$; $\mathbb{E}\{\cdot\}$ denotes the statistical expectation. 

\section{System Model} 
\subsection{CSI Feedback Model} 
We consider a FDD massive MIMO system with $N_{\text{T}}$ antennas, which serves a single-antenna UE. In order to acquire the downlink CSI knowledge for efficient precoding and beamforming, BS first sends pilot signals in the downlink channel for channel estimation. Then, the estimated CSI matrix at the user can be formulated as $\mathbf{H}=[\mathbf{h}_{1},\mathbf{h}_{2},\ldots,\mathbf{h}_{N_{\rm{c}}}]$, where $\mathbf{h}_{k}\in\mathbb{C}^{N_{\rm{T}}\times 1}$ denotes the CSI of $k$th subcarrier and $N_{\rm{c}}$ denotes the number of subcarriers. When each element of CSI matrix $\mathbf{H}$ is directly fed back to the BS, the length of feedback bits is proportional to the number of BS antennas, which poses an excessive overhead for the massive MIMO systems. Therefore, low-overhead CSI feedback is critical in FDD massive MIMO systems. 

The workflow of the deep learning-based CSI feedback is discussed as follows. In the UE side, the CSI matrix $\mathbf{H}$ is firstly transformed into the angular-delay domain $\widetilde{\mathbf{H}}$ with $\widetilde{\mathbf{H}}=\mathbf{F}_{\rm{a}}\mathbf{H}\mathbf{F}_{\rm{d}}^{H}$,
where $\mathbf{F}_{\rm a}\in\mathbb{C}^{N_{\rm{T}}\times N_{\rm{T}}}$ and $\mathbf{F}_{\rm{d}}\in\mathbb{C}^{N_{\rm{c}}\times N_{\rm{c}}}$ denote the normalized DFT matrices. Compared to the original CSI matrix $\mathbf{H}$, the transformed $\widetilde{\mathbf{H}}$ exhibits obvious sparsity in the angular-delay domain, which facilitates low-overhead feedback. Then, $\widetilde{\mathbf{H}}$ is fed into the encoder module to generate compressed codeword $\mathbf{c}=f_{\rm{en}}(\widetilde{\mathbf{H}})\in\mathbb{R}^{M\times 1}$, where $f_{\text{en}}(\cdot)$ denotes the learnable encoder module and $M\ll N_{\rm{T}}N_{\rm{c}}$ denotes the codeword length. Next, the compressed codeword is quantized into bits $\mathbf{b}=Q(\mathbf{c})$, where $Q(\cdot)$ denotes the quantization operation \cite{tvt_zhang_2024_quantization}. When BS receives feedback bits $\mathbf{b}$ via the uplink channel, it is input into the learnable decoder $f_{\text{de}}(\cdot)$ to reconstruct the channel. To optimize the learnable parameters in the encoder and decoder, mean square error (MSE) loss function 
\begin{equation}
    \label{equ: mse loss}
    \mathcal{L}=\Vert \widetilde{\mathbf{H}}-f_{\rm{de}}(Q(f_{\rm{en}}(\widetilde{\mathbf{H}})))\Vert_{F}^{2}
\end{equation}
is adopted during the offline training phase. 
\subsection{CSI Distribution Shift Model}
\label{subsec: distribution shift model}
\begin{figure*}[t]
        \centering
        \includegraphics[width=1\textwidth]{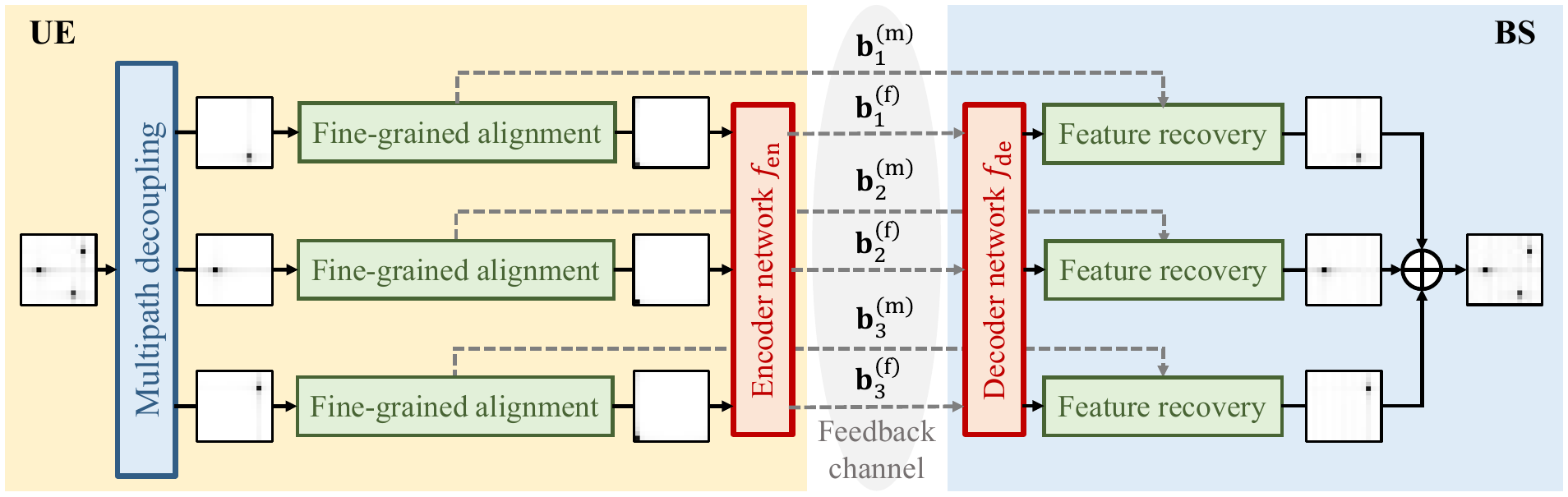}
    \captionsetup{font=footnotesize}
    \caption{Structure of proposed EG-CsiNet, where three path components are decoupled as an example. The amplitude of angular-delay representations for input, output, and intermediate path components are also illustrated.}
    \label{fig: model}
    \vspace{-15pt}
\end{figure*}
In order to increase the environment generalizability of the deep learning-based CSI feedback algorithms, the modeling of the distribution shift of CSI samples across the environments is crucial \cite{iclr_jivat_2015_modeling}. 
Based on the wideband geometrical channel model for massive MIMO system \cite{twc_he_2023_Beamspace}, channel $\mathbf{h}_{k}$ of $k$th subcarrier can be modeled as 
\begin{equation}
    \label{equ: channel model}
    \mathbf{h}_{k} = \sqrt{\frac{N_{\rm{T}}}{L}}\sum_{l=1}^{L}\alpha_{l}e^{-{\rm{j}}2\pi f_{k}\tau_{l}}\mathbf{a}(\phi_{l}),
\end{equation}
where $L$ denotes the number of paths; $f_{k}=f_{1}+(k-1)\Delta f$  denotes the frequency of $k$th subcarrier; $\alpha_{l}$, $\phi_{l}$, $\tau_{l}$ denote the complex gain, angle of departure (AoD), and delay of the $l$th physical path; $\mathbf{a}(\phi)$ denotes the steering vector of half-wavelength antenna array with
\begin{equation}
    \label{equ: array response}
    \mathbf{a}(\phi)=\frac{1}{\sqrt{N_{\rm T}}}\left[1, e^{{\rm j}\pi\sin{\phi}},\ldots,e^{{\rm j}(N_{\rm T}-1)\pi\sin{\phi}}\right]^{T}.
\end{equation}
Under fixed system parameters $\{N_{\rm T}, N_{\rm c}, \Delta f\}$ and array configuration in the massive MIMO system, the distribution of $\mathbf{H}$ is determined by the multupath parameters $\Theta=\{\Theta_{l}\}_{l=1}^{L}=\{\alpha_{l},\phi_{l},\tau_{l}\}_{l=1}^{L}$. When users are distributed in a specific environment, physical paths experience transmission, reflections, and diffractions with the interactions of the objects within the environment, which determine the marginal distribution of the parameters $\Theta_{l}$ and the distribution of the number of paths $L$. Meanwhile, parameters $\Theta_{l}$ of different paths are not independent but exhibit complex dependencies due to the relative layouts of different objects. Hereby, the number of paths and dependencies among paths can be merged as the multipath structure. Owing to the diverse user distributions and object layouts, both the distribution of multipath structure and single-path parameters obviously vary \cite{wang2025path}, which results in the significant distribution shift of CSI across environments. 

Further, the shift of marginal distribution of path parameters is investigated in the angular-delay domain. Define the response of $l$th path as $\mathbf{H}_{l}\in\mathbb{C}^{N_{\rm T}\times N_{\rm c}}$. Then, the $(m,n)$th element of the angular-delay domain representation $\widetilde{\mathbf{H}}_{l}=\mathbf{F}_{\rm a}\mathbf{H}_{l}\mathbf{F}_{\rm d}^{H}$ can be represented by 
\begin{equation}
    \label{equ: element}
    [\widetilde{\mathbf{H}}_{l}]_{m,n}\!=\!\frac{\alpha_{l}e^{{\rm j}\theta_{m,n}}}{\sqrt{L}}D_{N_{\rm{T}}}(i_{l}^{(\rm a)}\!+r_{l}^{(\rm a)}\!-\!m)D_{N_{\rm c}}(i_{l}^{(\rm d)}\!+r_{l}^{(\rm d)}\!-n),
\end{equation}
where function $D_{N}(x)=\frac{\sin(\pi x)}{\sin(\pi x/N)}$, angular-delay domain peak indexes $i_{l}^{(\rm a)}=\left[N_{\rm T}\sin(\phi_{l})/2\right]$, $i_{l}^{(\rm d)}=\left[N_{\rm c}\Delta f \tau_{l}\right]$, and the residues $r_{l}^{(\rm a)}=N_{\rm T}\sin(\phi_{l})/2-i_{l}^{(\rm a)}$, $r_{l}^{(\rm d)}=N_{\rm c}\Delta f\tau_{l}-i_{l}^{(\rm d)}$. Due to the limited number of antennas and subcarriers, residues $(r_{l}^{(\rm a)}, r_{l}^{(\rm d)})$ lead to the power leakage effect in the angular-delay domain \cite{tcom_ma_2021_deep}. For a given path, the peak positions and residues are determined by the interaction positions along the path, which are governed by the user positions and object layouts in the environments. Meanwhile, the materials of objects also impact the complex gain. Thus, distributions of peak indexes, power leakage, and complex gain of a single-path in the angular-delay domain drastically shift across environments. 


\section{EG-CsiNet: Environment-Generalizable Learning for CSI Feedback}
\subsection{Proposed EG-CsiNet}
\label{subsec: EG-CsiNet}

\subsubsection{Addressing Distribution Shift in CSI Feedback} Based on the CSI distribution shift model in Sec.~\ref{subsec: distribution shift model}, EG-CsiNet is proposed to enable environment-generalizable learning for CSI feedback, which is illustrated in Fig.~\ref{fig: model}. To address the distribution shift of multipath structure, the UE first applies multipath decoupling to the angular-delay domain representation $\widetilde{\mathbf{H}}$, which yields different path components. Then, encoder and decoder modules can individually compress and reconstruct each decoupled path component, where the distribution shift of the multipath structure can be removed. Further, fine-grained alignment is individually applied to the decoupled path components to address the single-path distribution shift. Specifically, the peak indexes of decoupled path components are aligned, and the distribution shifts of power leakage effect and complex gain are also mitigated. With the cascaded multipath decoupling and fine-grained alignment, the distribution shift of CSI across environments can be effectively addressed, which is interpretable in physics. 

\subsubsection{Multipath Decoupling} As shown in the leftmost of Fig.~\ref{fig: model}, the objective of multipath decoupling is to decompose the original angular-delay representation $\widetilde{\mathbf{H}}$ into a summation form, where the power distribution of each decoupled path component is concentrated in the angular-delay domain. For the angular-delay representations $\widetilde{\mathbf{H}}_{l}$, $\widetilde{\mathbf{H}}_{k}$ of $l$th and $k$th physical paths with distinct AoD and delay parameters, the row and column spaces of $\widetilde{\mathbf{H}}_{l}$ are approximately orthogonal to those of $\widetilde{\mathbf{H}}_{k}$. Additionally, the rank of the angular-delay representations $\widetilde{\mathbf{H}}_{l}$ is 1 for different physical paths. Inspired by the orthogonal property and rank-one property of different physical paths, a singular value decomposition (SVD)-based multipath decoupling in the angular-delay domain is introduced as follows. Firstly, $\widetilde{\mathbf{H}}$ is decomposed as $\widetilde{\mathbf{H}}=\sum_{i=1}^{R}\sigma_{i}\mathbf{u}_{i}\mathbf{v}_{i}^{H}$, where $\sigma_{i}$ denotes the $i$th largest sigular value, $\mathbf{u}_{i}$ and $\mathbf{v}_{i}$ denote the orthogonal bases. Then, the $i$th decoupled path component can be represented by $\widetilde{\mathbf{P}}_{i}=\sigma_{i}\mathbf{u}_{i}\mathbf{v}_{i}^{H}$. Since the encoder and decoder modules in EG-CsiNet operate in a path-wise manner, the feedback overhead is proportional to the number of decoupled path components. Therefore, to balance the decoupling performance and the feedback overhead, economical SVD is applied to determine the number of decoupled path components $\widehat{R}$. Explicitly, $\widehat{R}$ can be calculated by
\begin{equation}
    \label{equ: decoupled components}
    \widehat{R}=\min r, \quad \textnormal{s.t.}\sum_{i=1}^{r}\sigma_{i}^2\geq\eta\Vert\mathbf{H}\Vert_{F}^{2}, 
\end{equation}
where $\eta\leq1$ is a pre-defined threshold. Consider the sparse nature of $\widetilde{\mathbf{H}}$, a small amount of singular values $\{\sigma_{i}\}$ accounts for a large portion of the channel power, which also facilitates the low feedback overhead. 

\subsubsection{Fine-grained Alignment} The objective of fine-grained alignment is to address the distribution shift of the decoupled path components. For the simplicity of expression, subscript $i$ in $\widetilde{\mathbf{P}}_{i}$ is omitted and the spatial-frequency domain representation of $\widetilde{\mathbf{P}}$ can be calculated $\mathbf{P}=\mathbf{F}_{\rm a}^{H}\widetilde{\mathbf{P}}\mathbf{F}_{\rm d}$. To simultaneously address the distribution shift of peak indexes and mitigate the power leakage effect, the oversampled DFT codebook can be adopted to scan the fine-grained peak position in both the angular and delay domains. Explicitly, the $n$th codeword for the angular-domain $O_{\rm a}$-oversampled DFT codebook can be defined as 
\begin{equation}
    \label{equ: angular codeword}
    \mathbf{w}_{n}^{(\rm a)}=\left[1,e^{{\rm j}2\pi \frac{n}{O_{\rm a}N_{\rm T}}},\ldots,e^{{\rm j}2\pi \frac{n(N_{\rm T}-1)}{O_{\rm a}N_{\rm T}}}\right]^{T},
\end{equation}
where $0\leq n\leq O_{\rm a}N_{\rm T}-1$. Then, the fine-grained peak position in the angular-domain can be calculated by 
\begin{equation}
    \label{equ: peak position angular}
    n^\star=\mathop{\arg\max}_{n}\left\{\Vert(\mathbf{w}_{n}^{(\rm a)})^{H}\mathbf{P}\Vert_{2}^{2}\right\}. 
\end{equation}
Similarly, the $m$th codeword for the delay-domain $O_{\rm d}$-ovresampled DFT codebook can be formulated as
\begin{equation}
    \label{equ: delay codeword}
    \mathbf{w}_{m}^{(\rm d)}=\left[1,e^{{\rm j}2\pi \frac{m}{O_{\rm d}N_{\rm c}}},\ldots,e^{{\rm j}2\pi \frac{m(N_{\rm c}-1)}{O_{\rm d}N_{\rm c}}}\right]^{T},
\end{equation}
and the fine-grained delay-domain peak position is yielded by
\begin{equation}
    \label{equ: peak position delay}
    m^\star=\mathop{\arg\max}_{m}\left\{\Vert\mathbf{P}\mathbf{w}_{m}^{(\rm d)}\Vert_{2}^{2}\right\}. 
\end{equation}
To align the peak value to a fixed position in the angular-delay domain, phase of each element in $\mathbf{P}$ can be adjusted accordingly, where the phase adjustment matrix $\mathbf{S}\in\mathbb{C}^{N_{\rm T}\times N_{\rm c}}$ can be calculated by 
\begin{equation}
    \label{equ: phase adjust}
    \mathbf{S}=\text{conj}(\mathbf{w}^{(\rm a)}_{n^\star})\otimes(\mathbf{w}_{m^\star}^{(\rm d)})^{T}.
\end{equation}
Based on the scanned fine-grained positions $(n^\star,m^\star)$, we can calculate the peak value of $\mathbf{P}$ by $p=(\mathbf{w}_{n^\star}^{(\rm a)})^{H}\mathbf{P}\mathbf{w}_{m^\star}^{(\rm d)}$. With the multipath decoupling step, peak value $p$ can largely reflect the complex path gain. Therefore, we can quantize the phase of peak value $p$ with $Q_{\rm p}$-bit uniform quantization, which yields $\beta=Q(\angle p)$. Based on the aforementioned processing, the aligned path component $\widetilde{\mathbf{P}}^{(\rm aln)}$ in the angular-delay domain is yielded by
\begin{equation}
    \label{equ: align}
    \widetilde{\mathbf{P}}^{(\rm aln)}=\mathbf{F}_{\rm a}(e^{-\rm{j}\beta}\mathbf{S}\odot\mathbf{P})\mathbf{F}_{\rm d}^{H},
\end{equation}
which can effectively address the distribution shift of single-path across environments, including peak indexes, power leakage, and complex gain. Once the fine-grained alignment is applied, the metadata $(n^\star,m^\star,\beta)$ of each aligned path component is also fed back to the BS during the model inference phase, which is detailed in \ref{subsec: training}. 
\subsection{Training, Inference and Feedback Overhead of EG-CsiNet}
\label{subsec: training}
\subsubsection{Model Training} In the offline model training phase, an aligned path component training dataset should be first yielded based on the multipath decoupling and fine-grained alignment steps in Sec.~\ref{subsec: EG-CsiNet}. Note that the $\widetilde{\mathbf{P}}^{(\rm aln)}$ has the same shape $N_{\rm T}\times N_{\rm c}$ with $\widetilde{\mathbf{H}}$ and both $\widetilde{\mathbf{P}}^{(\rm aln)}$ and $\widetilde{\mathbf{H}}$ exhibit sparse nature in the angular-delay domain. Thus, the NN of the encoder and decoder module in the EG-CsiNet can adopt the same structure as the conventional deep learning-based CSI feedback algorithms, including the CNN \cite{wcl_wen_2018_csinet,twc_guo_2020_csinetp} and transformer \cite{wcl_cui_2022_transnet}. Consistent with the multipath-decoupling and fine-grained alignment, the encoder module and decoder module are trained with the aligned path component, where the training loss is defined as 
\begin{equation}
    \label{equ: aligned loss}
    \mathcal{L}_{\text{EG-CsiNet}}=\Vert\widetilde{\mathbf{P}}^{(\rm aln)}-f_{\rm de}(Q(f_{\rm en}(\widetilde{\mathbf{P}}^{(\rm aln)})))\Vert_{F}^{2}.
\end{equation}
Thus, the optimal parameters in $f_{\rm en}(\cdot)$ and $f_{\rm de}(\cdot)$ are yielded by minimizing $\mathcal{L}_{\text{EG-CsiNet}}$, which are retained in the model inference phase after deployment. 

\subsubsection{Model Inference} In the online model inference phase, the reconstructed CSI matrix that composes multiple path components is required. As shown in Fig.~\ref{fig: model}, additional metadata of each aligned path component is also fed back to the BS, including the oversampled peak positions $(n^\star,m^\star)$ and quantized peak phase $\beta$. Define the output of the decoder module as $\widehat{\widetilde{\mathbf{P}}}^{(\rm aln)}=f_{\rm de}(Q(f_{\rm en}(\widetilde{\mathbf{P}}^{(\rm aln)}))$. Then, $\widehat{\widetilde{\mathbf{P}}}^{(\rm aln)}$ and the related metadata $(n^\star,m^\star,\beta)$ is fed into the feature recovery module, which can relcoate the peak position in the angular-delay domain to $(n^\star,m^\star)$ and recover the peak phase. Explicitly, the recovered path component $\widehat{\mathbf{P}}\in\mathbb{C}^{N_{\rm T}\times N_{\rm c}}$ in the spatial-frequency domain can be represented by 
\begin{equation}
    \label{equ: recovery}
    \widehat{\mathbf{P}}=\text{conj}(e^{-{\rm j}\beta}\mathbf{S})\odot\left(\mathbf{F}_{\rm a}^{H}\widehat{\widetilde{\mathbf{P}}}^{(\rm aln)}\mathbf{F}_{\rm d}\right).
\end{equation}
Denote the $i$th recovered path component as $\widehat{\mathbf{P}}_{i}$. Then, the reconstructed multipath channel $\widehat{\mathbf{H}}$ in the spatial-frequency domain is yielded by summing up all recovered path components, i.e.,
\begin{equation}
    \label{equ: reconstruct channel}
    \widehat{\mathbf{H}}=\sum_{i=1}^{\widehat{R}}\widehat{\mathbf{P}}_{i}.
\end{equation}
\subsubsection{Feedback Overhead} As shown in the middle of Fig.~\ref{fig: model}, the feedback bits for $i$th decoupled path component include two parts, the compression bits $\mathbf{b}^{(\rm f)}_{i}$ for aligned path component and the related metadata bits $\mathbf{b}^{(\rm m)}_{i}$. Based on the size of oversampled DFT codebooks, the length of metadata bits is $q_{\rm m}=Q_{\rm p}+\lceil\log_{2}(O_{\rm a}N_{\rm T}O_{\rm d}N_{\rm c})\rceil$ for each decoupled path. Assume $Q_{\rm f}$-bit element-wise quantization is applied in the encoder module, and the feedback bits for compressing each aligned path component is $q_{\rm f}=MQ_{\rm f}$. Based on the path-wise feedback manner in EG-CsiNet, the length of total feedback bits for a single channel instance is $q=\widehat{R}(q_{\rm m}+q_{\rm f})$. For the users distributed in a specific environment, the number of decoupled path components $\widehat{R}$ varies with the user location. For instance, when the user is located at the line-of-sight (LOS) areas, the LOS path accounts for a large portion of the channel power, and $\widehat{R}$ is usually limited. Conversely, when the user is located in the non-line-of-sight (NLOS) areas, the channel composes multiple paths under reflections, diffractions, and scatterings with distinct AoDs and delays. Then, power distribution of the channel in angular-delay domain is more dispersive and $\widehat{R}$ is relatively high. Since the feedback overhead is proportional to $\widehat{R}$, the proposed EG-CsiNet can adaptively adjust the feedback overheard for each channel instance, which statistically achieves low-overhead CSI feedback. Thus, in a specific environment, the average feedback overhead for EG-CsiNet is $\mathbb{E}\{\widehat{R}\}(q_{\rm m}+q_{\rm f})$. 
\section{Simulation Study}
\subsection{Simulation Setup}

In the simulation, the open-source dataset WAIR-D \cite{huangfu2022wair} is adopted to generate the CSI data, which is built on realistic maps from major cities worldwide and involves diverse building layouts. Examples of building layouts can also be found in the bottom of Fig.~\ref{fig: generalizability}. Thus, the distribution of CSI samples drastically shifts across different environments. 
The massive MIMO system operating at 2.6 GHz is deployed in different environments. A uniform planar array (UPA) with $N_{\rm T}=32$ antennas is equipped in the BS, where the number of horizontal and vertical antennas is set as 8 and 4, respectively. The number of UE antenna is set as 1. System bandwidth is set as 10 MHz, where the number of subcarrier is set as $N_{\rm c}=32$.


The configurations for the proposed EG-CsiNet and the baselines are detailed as follows. Vanilla AE and UniversalNet+ \cite{wcnc_liu_2024_generalization} are adopted as the deep learning-based baselines. The vanilla AE refers to standard end-to-end training of the encoder and decoder without specialized generalization components, which includes single-environment training and multiple environment mixing \cite{twc_jiang_2024_multidomain}. Additionally, the enhanced type-II (eType-II) codebook \cite{3gpp.38.214} is adopted as a baseline without deep learning. For the NN in the encoder and decoder modules of the deep learning-based baselines and EG-CsiNet, the standard CsiNet structure \cite{wcl_wen_2018_csinet} is adopted by default, and the quantization bit for the codeword is set as $q_{\rm f}=6$ bit. In each training environment, the number of training samples is set as 9000. During the training, the Adam optimizer with initial learning rate $10^{-3}$ is adopted, and the batch size is set as 64. To justify the environment generalizability, the trained model is tested in the 90 unseen environments with a total of 90000 CSI samples, where the normalized mean square error $\text{NMSE}=\mathbb{E}\{\Vert\widehat{\mathbf{H}}-\mathbf{H}\Vert_{F}^{2}/{\Vert\mathbf{H}\Vert_{F}^{2}}\}$
is adopted as the performance metric. In the proposed EG-CsiNet, the threshold in multipath decoupling is set as $\eta=0.99$, and the oversampling factors in horizontal, vertical, and delay domains are set as 2. The peak phase quantization bit is set as $Q_{\rm p}=2$ bit. The performances of deep learning models are averaged over 3 random initializations.

\subsection{Simulation Results}
\begin{figure}[t]
        \centering
        \includegraphics[width=0.45\textwidth]{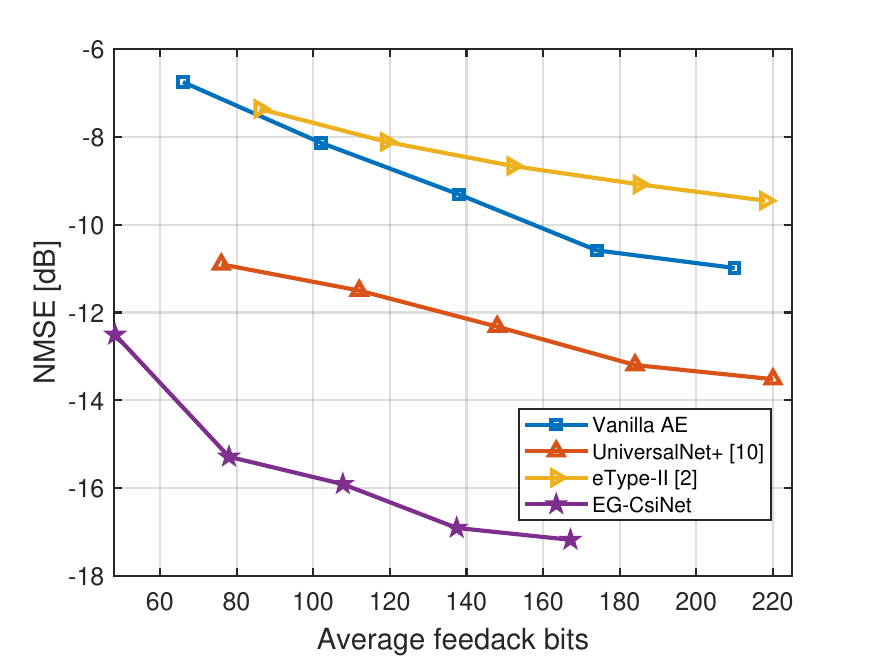}
    \captionsetup{font=footnotesize}
    \caption{Intra-environment NMSE of CSI feedback schemes with varying feedback bits. }
    \label{fig: intra-1}
    \vspace{-15pt}
\end{figure}

\subsubsection{Intra-Environment Validation} Before the investigation of environment-generalizability, the intra-environment performance of EG-CsiNet is verified. Hereby, we consider a single training environment, and the NMSE in the same environment with different feedback bits is shown in Fig.~\ref{fig: intra-1}. Compared to the baselines, the proposed EG-CsiNet can reduce NMSE for around 4.5 dB in the training environment with the same feedback overhead level. Therefore, the proposed EG-CsiNet can also realize efficient intra-environment CSI feedback when the training samples from the target environment are available.

\subsubsection{Environment-Generalizability Comparison}

\begin{figure}[t]
        \centering
        \includegraphics[width=0.45\textwidth]{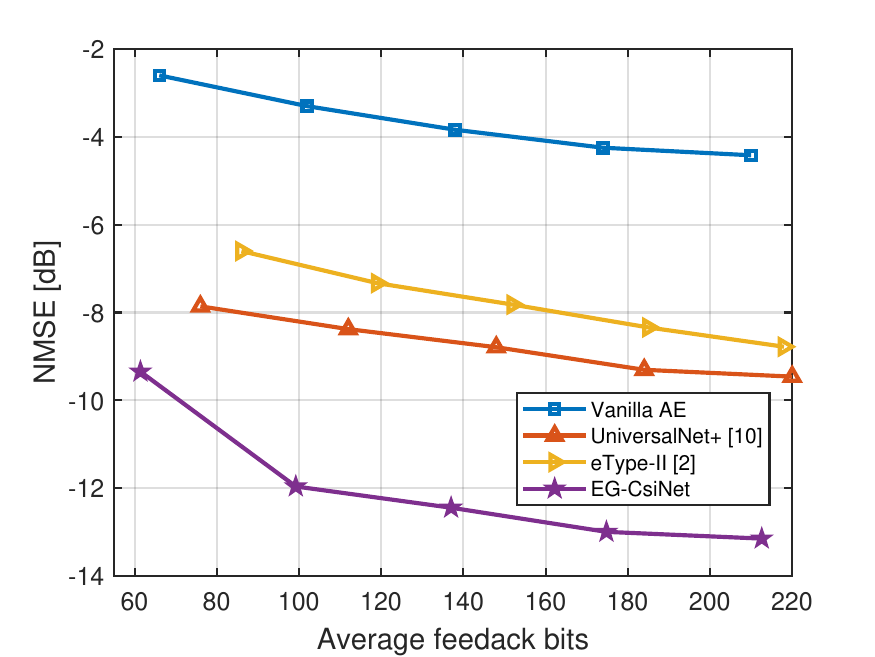}
    \captionsetup{font=footnotesize}
    \caption{NMSE of CSI feedback schemes in 90 unseen environments with varying feedback bits, where the number of training environment is set as 1. }
    \label{fig: generalizability-1}
    \vspace{-15pt}
\end{figure}

\begin{figure}[t]
        \centering
        \includegraphics[width=0.45\textwidth]{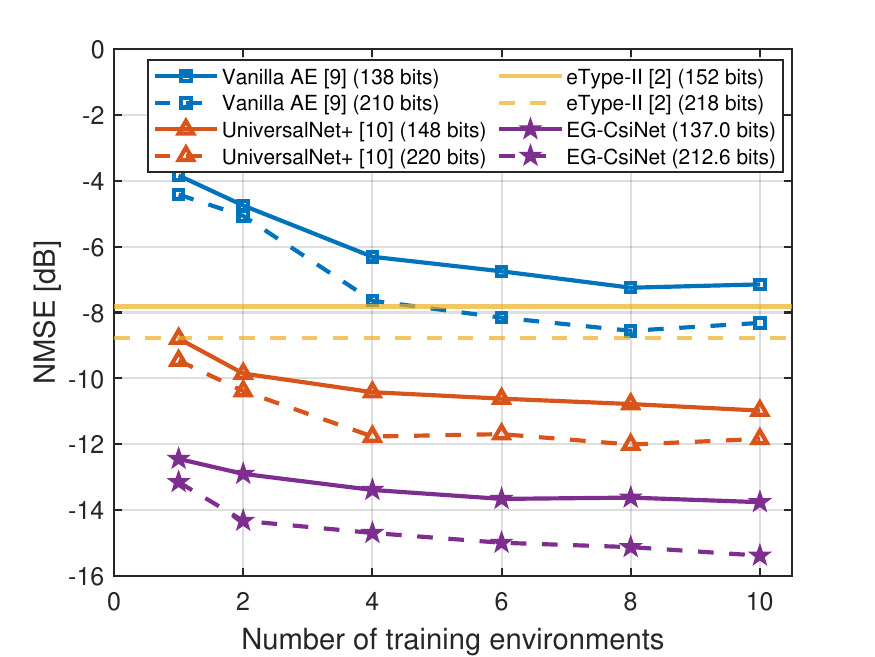}
    \captionsetup{font=footnotesize}
    \caption{NMSE of CSI feedback schemes in 90 unseen environments with varying training environments.}
    \label{fig: generalizability-3}
    \vspace{-10pt}
\end{figure}

\begin{figure}[t]
    \centering
    \subfloat[Pretrained in 1 environment]
    {\includegraphics[width=0.44\textwidth]{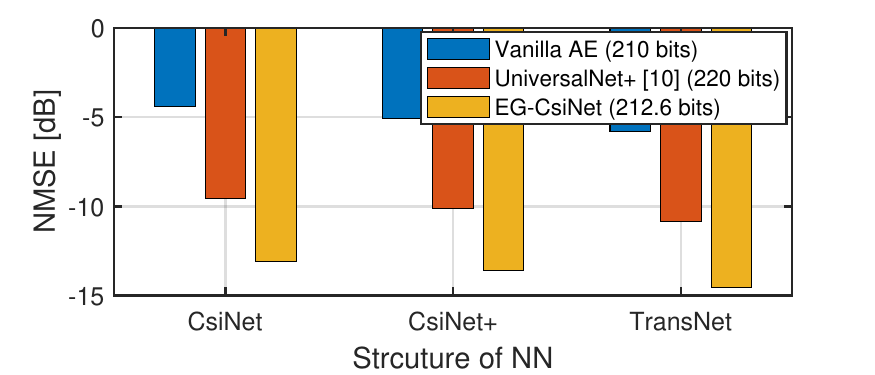}\label{subfig: impact of vd}}\quad
    \subfloat[Pretrained in 10 environments]
    {\includegraphics[width=0.44\textwidth]{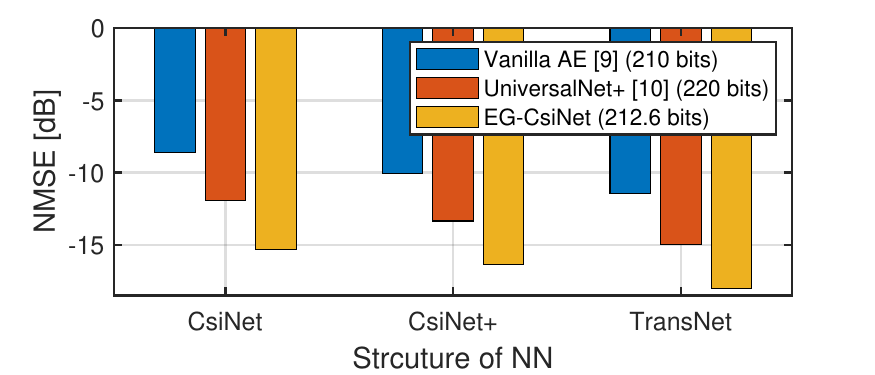}\label{subfig: impact of vw}}
    \captionsetup{font=footnotesize}
    \caption{Environment generalizability comparison with different NN structures.}
    \label{fig: NN structure}
    \vspace{-15pt}
\end{figure}

Firstly, we set the number of the training environment as 1 and investigate the NMSE under different average lengths of feedback bits, which is plotted in Fig.~\ref{fig: generalizability-1}. Firstly, it can be found that the NMSE of the proposed EG-CsiNet has been reduced by more than 3.5 dB compared to the SOTA in unseen environments. Compared to UniversalNet+, the proposed EG-CsiNet can better address the distribution shift of CSI samples across different environments, especially in the NLOS scenarios with complex multipath propagation. Additionally, the learnable encoder and decoder modules in EG-CsiNet can better capture the correlations in the angular-delay domain of the aligned path components, which exhibits better compression capability than the eType-II codebook with the guarantee of environment-generalizability. On the contrary, due to the drastic CSI distribution shift, the vanilla AE cannot achieve accurate channel feedback in unseen environments with a single source. 


Then, the environment generalizability with a varying number of training environments is investigated, which is depicted in Fig.~\ref{fig: generalizability-3}. It can be found that the proposed EG-CsiNet still achieves the best feedback performance in unseen environments with multiple training environments compared to the SOTA. Additionally, the environment-generalizability of the proposed EG-CsiNet can also be gradually improved with an increasing number of training environments, which also facilitates its deployment under different availability of training data sources. 

Further, environment-generalizability comparison with different NN structures is depicted in Fig.~\ref{fig: NN structure}, which includes the CNN-based CsiNet/CsiNet+ and the transformer-based TransNet. It can be found that the proposed EG-CsiNet can achieve the best generalizability with different types of NN, which is compatible with various NN structures. Additionally, it can be found that the EG-CsiNet with a small-sized CsiNet structure can even achieve lower NMSE compared to the baselines with a large-sized TransNet structure, which also facilitates the deployment with limited memory resources.

\section{Conclusion}
In this paper, environment-generalizable learning for CSI feedback is realized with explicit physics interpretation. Firstly, we rigorously model the distribution shift of CSI across environments based on the propagation of paths in the wireless channel, which is composed of the distribution shift of multipath structure and a single-path. Then, EG-CsiNet is proposed to enhance the deep learning-based CSI feedback to address the distribution shift of CSI. In EG-CsiNet, a SVD-based multipath decoupling module is introduced to individually feed back each decoupled path component, which can address the distribution shift of multipath structure and enable overhead-adaptive feedback. Then, fine-grained alignment is designed to address the distribution shift of a single-path, where the corresponding feature recovery module in the model inference phase is also implemented. Comprehensive simulations indicate that the EG-CsiNet can greatly enhance the generalizability to unseen environments compared to the SOTA. 
\section{Acknowledgement}
This work was supported by the National Key R\&D Program of China under Grant 2022YFB2902004.
\bibliographystyle{IEEEtran}
\bibliography{Ref}{}
\end{document}